# Simulation of conduction cooling of a 650 MHz 5-cell cavity

Roman Kostin

Euclid Techlabs, LLC., Bolingbrook, IL 60440

## Abstract

The research note presents results of coupled RF and thermal simulation of a cryocooler conduction cooled 650 MHz SRF cavity made of bulk niobium and coated with Nb$_3$Sn on the RF surface. The cavity is part of a particle accelerator design capable of producing 10 MeV, 1000 kW electron beam for application in wastewater treatment. More details of the accelerator design are presented in R.C. Dhuley *et al.* (2022) Phys. Rev. Accel. Beams 25, 041601.

## 1. Introduction

Cryocooler conduction cooling of superconducting radiofrequency (SRF) Nb$_3$Sn cavities [1] is a dramatically simpler cooling scheme than the conventional liquid helium bath cooling. The simpler and more reliable cryogenic scheme [2-7] can lead to wider adoption of the SRF accelerator technology, particularly in the industrial area where high power electron beams are required. Some examples are irradiation treatment of industrial and municipal wastewater, medical device sterilization, asphalt pavement curing, and scientific instruments such as UED and UEM [8-10].

Due to limited cooling capacity of present day cryocoolers, it is imperative to properly design the conduction cooling scheme than enhances the total thermal conductance of the cryocooler-cavity system and also designing a cryostat enclosure that reduces the total heat leak into the system.

The cryostat design is presented in [8] while this research note focuses on thermal simulation results of the conduction cooling system.

## 2. Simulation inputs

Thermal simulations of a 5-cell 650MHz $Nb_3Sn$ cavity is performed. The cavity provides 10 MeV beam that is accelerated from 300 keV by external injection. The 3D model of the cavity is presented in Fig. 1. As one can see from this figure, six PT-420 are used for the cavity cells cooling and two PT-425 are used for intercepting heat from RF fundamental power couplers. Fig. 2 shows cooling capacity of the two cryocoolers.

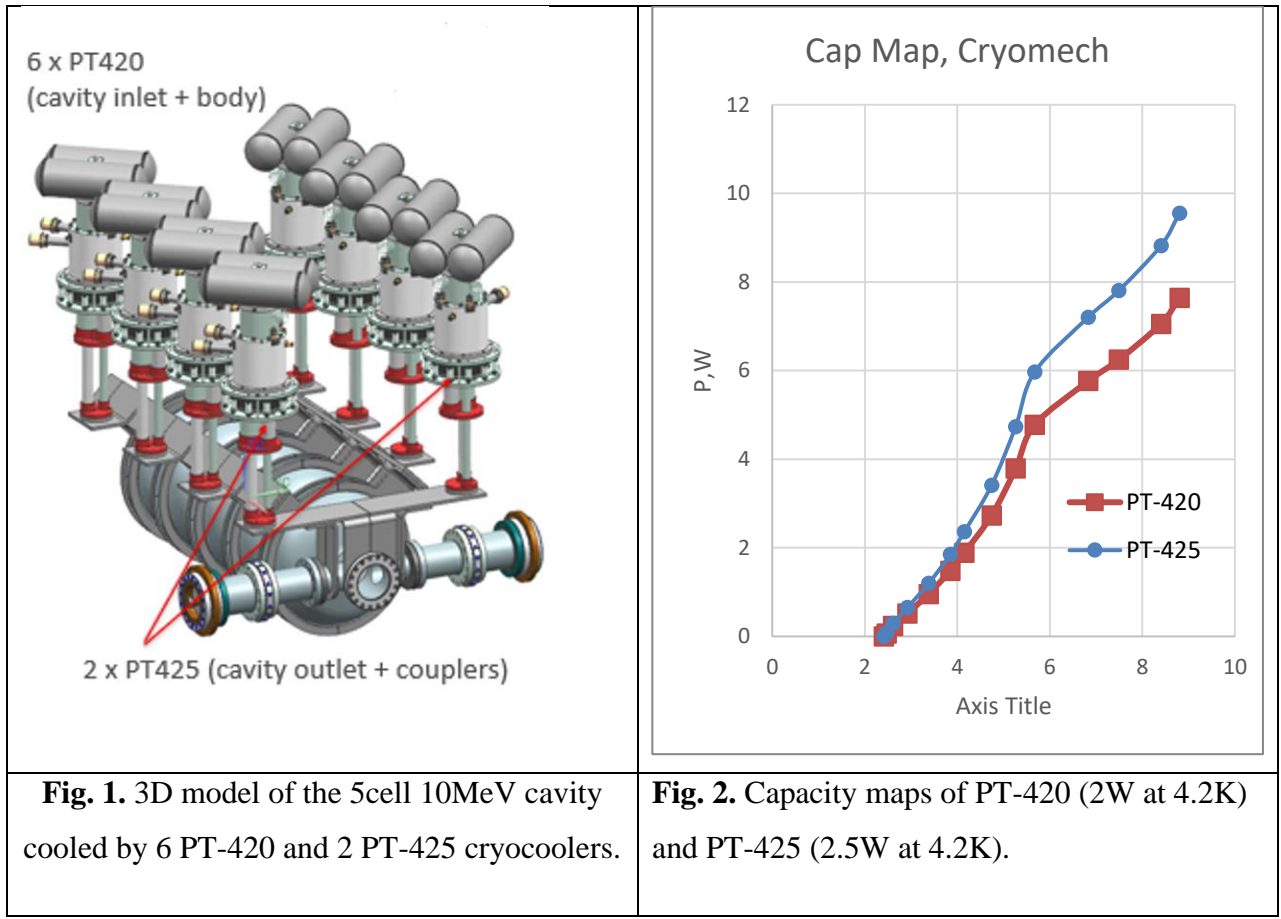

**Fig. 1.** 3D model of the 5cell 10MeV cavity cooled by 6 PT-420 and 2 PT-425 cryocoolers.

**Fig. 2.** Capacity maps of PT-420 (2W at 4.2K) and PT-425 (2.5W at 4.2K).

The heat sources to the cavity are presented in Table 1. These are conduction to the cavity beampipes, thermal radiation, dynamic losses from RF and some heat dissipation due to the beam loss. The total heat dissipation is around 20 W at 5 K for the cavity, scaled from a quality factor of $Q_0=2e10$ at 4.4 K.

Material properties required for the simulation are taken from the following sources: thermal conductivity of bulk niobium from [11], BCS resistance of Nb$_3$Sn from [12], thermal conductivity of bulk aluminum from [13], thermal contact resistance from [14,15], and cryocooler capacity from [16].

Table 1: Expected 5 K heat load on the 5-cell 650 MHz SRF cavity

| Component | | Heat load [W] | Comment |
|---|---|---|---|
| Cavity body | Dynamic | 11.7 | RF dissipation at 5 K |
| | Radiation | 0.05 | from thermal shield |
| | Beam loss | 1 | 1e-6 loss |
| | Conduction | 0.1 | from supports |
| | Total | 12.9 | |
| Cavity injection side | Conduction | 0.05 | Beam pipe conduction |
| | Radiation | 0.24 | Input port radiation |
| | Total | 0.29 | |
| **Cavity body + injection side** | | **13.2** | |
| | | | |
| Cavity high-energy side | Coupler | 6 | two couplers, static + dynamic, scaled from 100 kW |
| | Conduction | 0.05 | Beam pipe conduction |
| | Radiation | 0.24 | Input port radiation |
| | Total | 6.3 | |
| **Cavity high-energy side** | | **6.3** | |
| | | | |
| **Cavity total** | | **19.5** | |

There are two different types of cavities: with internal and external injection. The cavity with internal injection has lower energy beampipe opening of 35 mm and with external – 100 mm. Both cavities' electrodynamic paramters are presented in Table 2.

Table 2: "Preliminary" cavity parameters at 10 MV voltage gain.

| Scale | 0.7, Ø100 | 0.7, Ø35 |
|---|---|---|
| $R/Q$, Ω | 656.3 | 657.3 |
| $G$, Ω | 259.5 | 260.5 |
| $R/Q*G$, Ω² | 170310 | 171230 |
| $E_{s,max}$, MV/m | 21.2 | 21. |
| $H_{s,max}$, mT | 34.8 | 34.3 |
| $P_{loss}$, W | $0.6 \cdot R_s$ | $0.6 \cdot R_s$ |

It is worth to mention, that as was found later, surface electric field for 10 MeV energy gain correpsonds to 36.5 mT and 17.5 MV/m [8]. The cavity under imvetigation is the cavity with external injection. The provided information above is enough to proceed with thermal simulations. We start with RF simulations to generate dynamic losses for the thermal module but first field scaling to the required level is neededSimulations

## 3. 1st geometry case: Smaller OD beampipe

### a. RF simulation and particle tracking

RF simulations were performed in order to provide a dynamic heat load for the cavity thermal simulations. RF fields scaling to the required level is needed, i.e. when the cavity can provide an energy gain of 10 MeV. The results obtained in Comsol were scaled by magnetic field to 36.5 mT which corresponds to 17.5 MV/m of electric field on surface and 18 MV/m electric field amplitude on axis. One can find field distributions on axis and on the cavity surface on Fig.3. Single particle tracking was done to find out if the obtained level of fields is enough to provide 10 MeV gain. The results of single particle tracking is presented on Fig. 4.

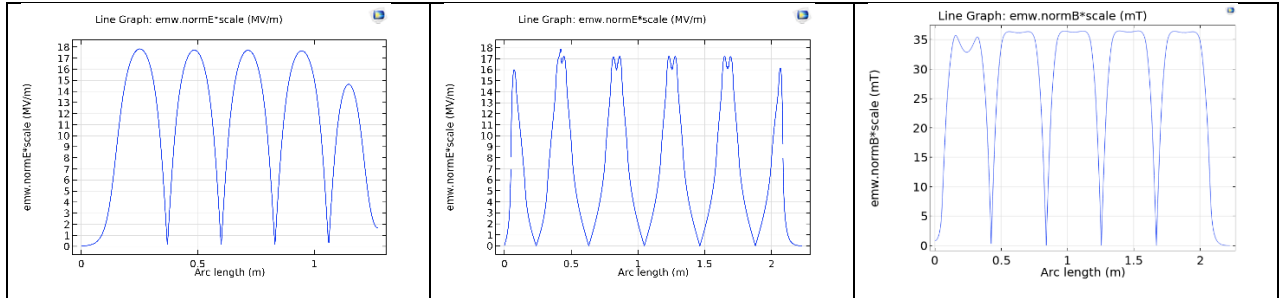

| Electric field on the axis | Electric field on the surface | Magnetic field on the surface |

**Fig. 3.** Electric field and Magnetic field in the 650 MHz 5-cell cavity with external injection to provide 10 MeV gain to the beam.

The beam tracking was performed for the case of 18 MV/m on the axis, which corresponds to 36.5 mT peak magnetic field on the surface. The cavity provides 10 MeV energy gain for this field level. The maximm energy gain is at -15 deg and corresponds to 10.35 MeV.

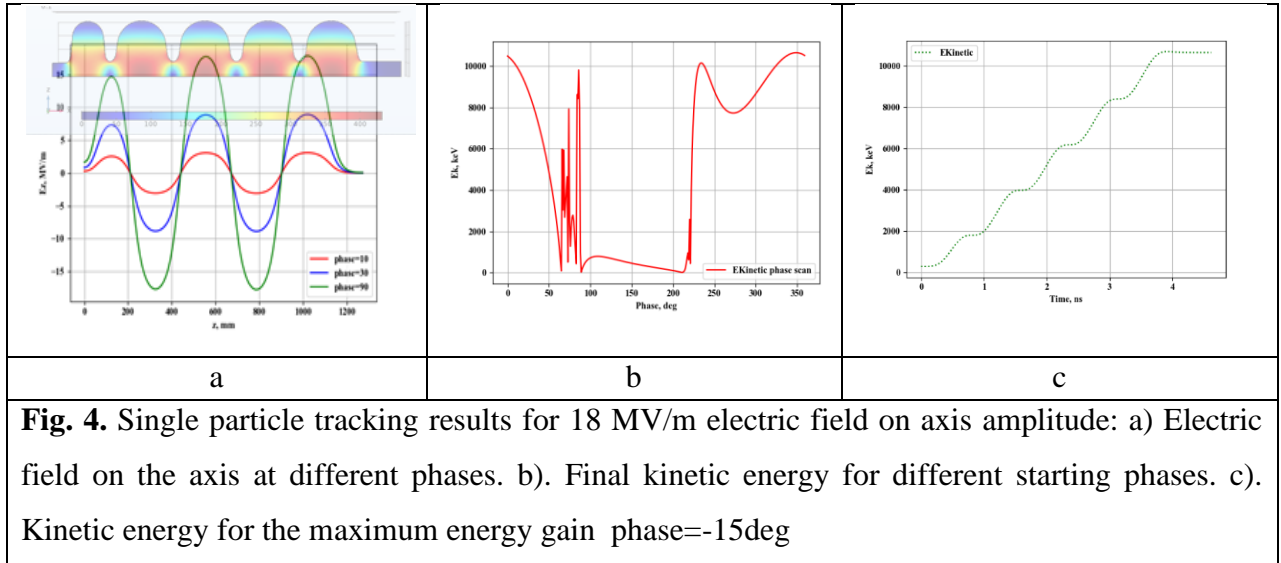

| a | b | c |

**Fig. 4.** Single particle tracking results for 18 MV/m electric field on axis amplitude: a) Electric field on the axis at different phases. b). Final kinetic energy for different starting phases. c). Kinetic energy for the maximum energy gain phase=-15deg

As a conclusion for this section, 18 MV/m peak on axis field in the cavity is enough to provide energy gain of 10 MeV. Peak magnetic field on the surface is 36.5 mT, peak electric field on surface is 17.5 MV/m. This field level will be used for thermal simulation.

## b. Thermal simulations, $Q_0$=1.5e10 (Rs=10 nOhm).

Thermal simulation of the cavity is required which has temperatures of PT-420 and PT-425 cryocoolers of T_pt420=4.4K and T_pt425=4.6K correspondingly. Material properties used in thermal simulations are plotted with temperature in Fig 5.

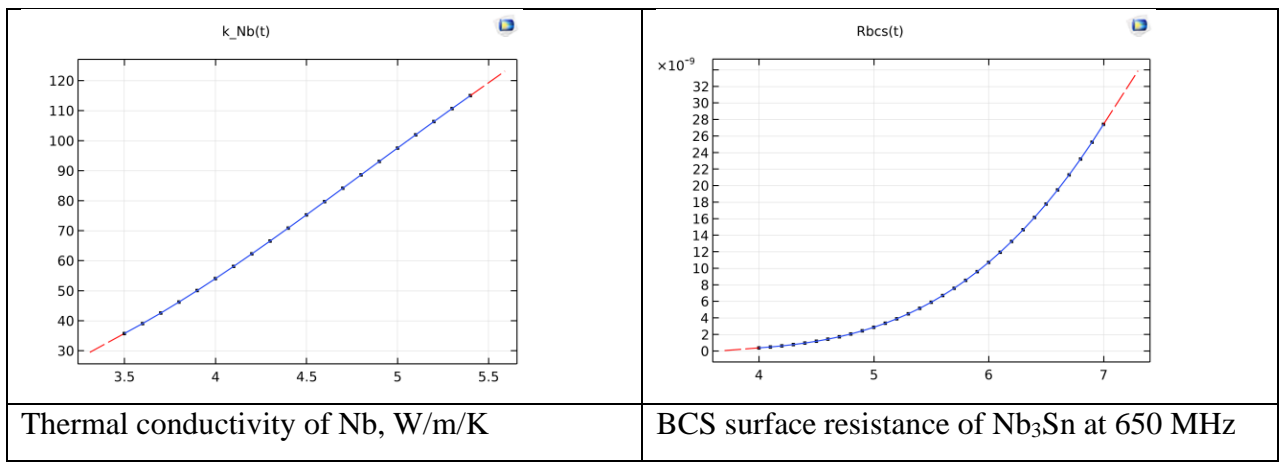

| Thermal conductivity of Nb, W/m/K | BCS surface resistance of $Nb_3Sn$ at 650 MHz |

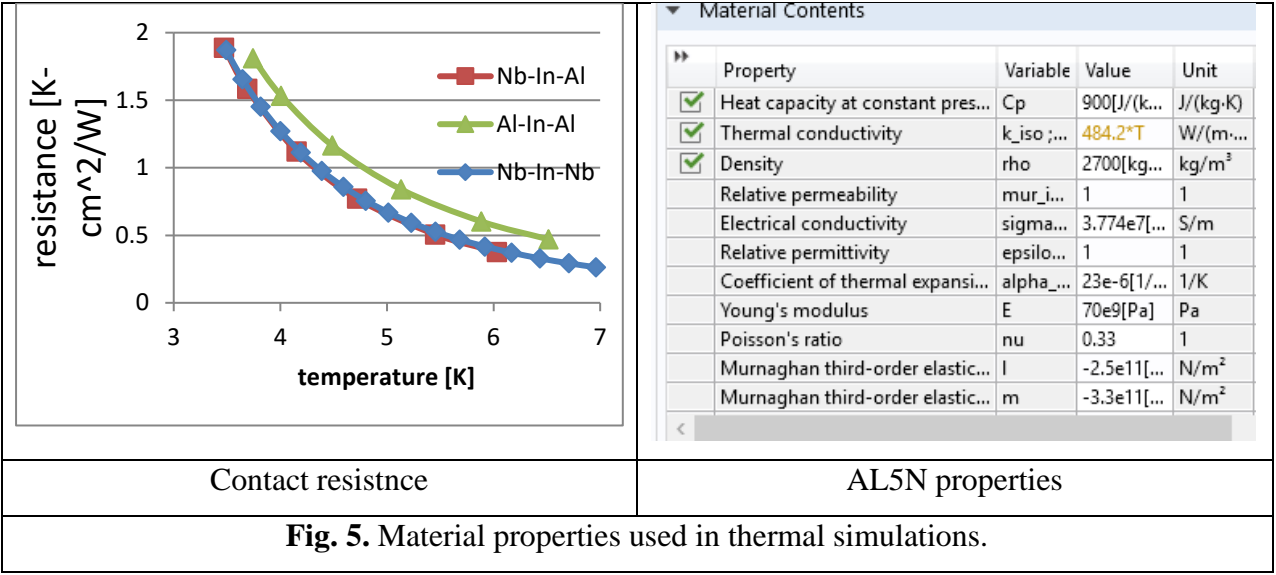

| Contact resistnce | AL5N properties |

**Fig. 5.** Material properties used in thermal simulations.

The static loads were applied according to the data provided in Table 1. Thermal contact resistance is applied between alluminum link and niobium ribs (Al-Nb), between alluminum links (Al-Al) and for niobium weld (Nb-Nb). The effect of contact resitsnace was investigated by gradually adding contact resitstance between different mating surfaces. Temperature distribution is presented below in Fig. 6 for the case of surface resistance of Rs=10 nOhm, which corresponds to the quality factor of $Q_0$=2.6e10.

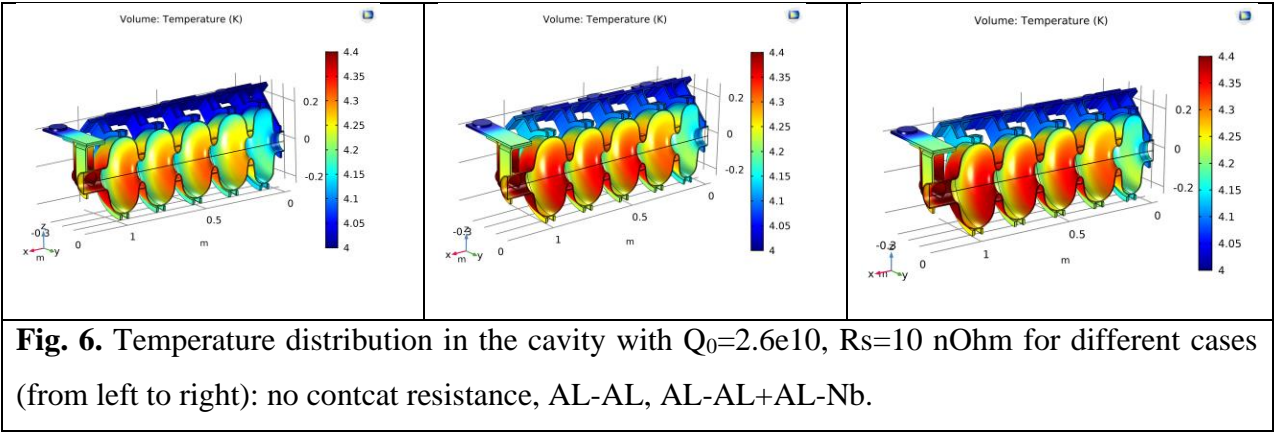

**Fig. 6.** Temperature distribution in the cavity with $Q_0$=2.6e10, Rs=10 nOhm for different cases (from left to right): no contcat resistance, AL-AL, AL-AL+AL-Nb.

As one can see from Fig. 6, the effect of contact resistance is clearly seen (note the color change with the same scale for all three cases): temperature of the cavity is increasing with more contact resistance added to the mating surfaces, however, it is fairly small – only 50 mK temperature rise.

The simulation of the cavity with $Q_0$=1.5e10 which corresponds to surface resistance of Rs=17.5 nOhm is presented on Fig. 7.

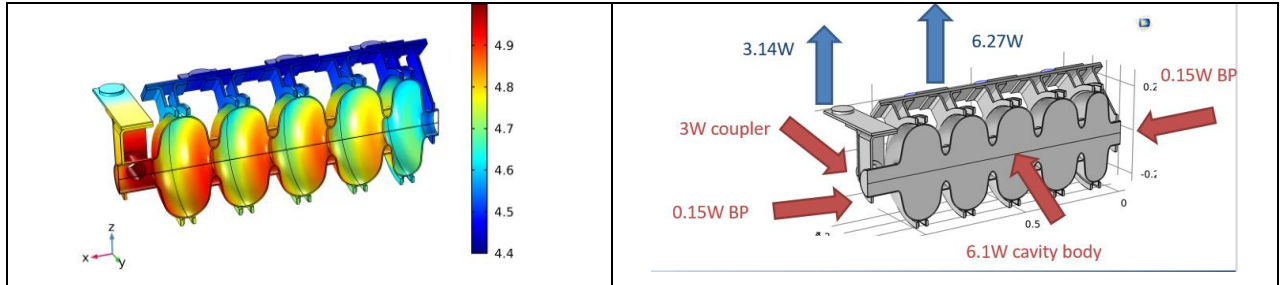

**Fig. 7.** Thermal simulation results for the 5-cell cavity with $Q_0$=1.5e10 (Rs=17.5 nOhm): temerature distribution (on the left) and power balance (on the right).

Power balance is presented in the Fig.7 on the right. The heat applied to the cavity is represented by red arrows, and the heat taken by the cryoccolers is represented by blue arrows. Power coming to the cavity is equal to the power taken by cryocoolers. The total capacity of cryocoolers at the given temperatures is higher than the power taken by cryoccolers, which means the cavity will stabilize at lower temperature and the current solution gives some safety margin to accommodate additional power dissipated in the cavity.

## c. Thermal simulations, $Q_0$=8e9 (Rs=32 nOhm)

Another case of thermal simulation includes the cavity with quality factor of $Q_0$=8e9, which corresponds to surface resistance of Rs=32 nOhm. The dynamic losses in this case should be around 22 W. The four PT-420 cryocoolers should have the temperature of T=5.35 K to accommodate this cavity dissipation according to the capacity map (see Fig. 2). At the same time, PT-425 will stabilize at 4.84 K. Both cases are depicted in Fig. 8.

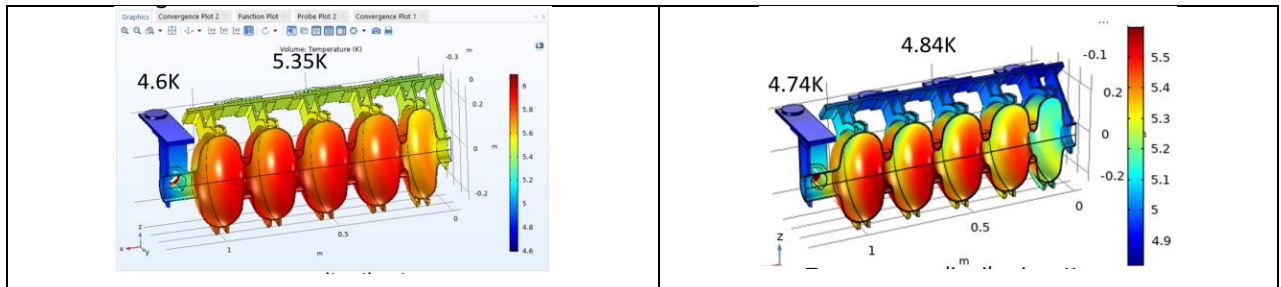

| PT420 and PT425 | PT425 only |

**Fig. 8.** Temperature distribution in the 5 cell cavity with Q0=8E9, cooled by PT-420, PT-425 on the left and PT-425 only on the right.

The latter case (PT425 only) was analysed and it was found that the power flows to the cryoccolers responsible for cells cooling are actually not the same because the temperature along the cavity is not uniform. The power balance can be found in the Fig. 9.

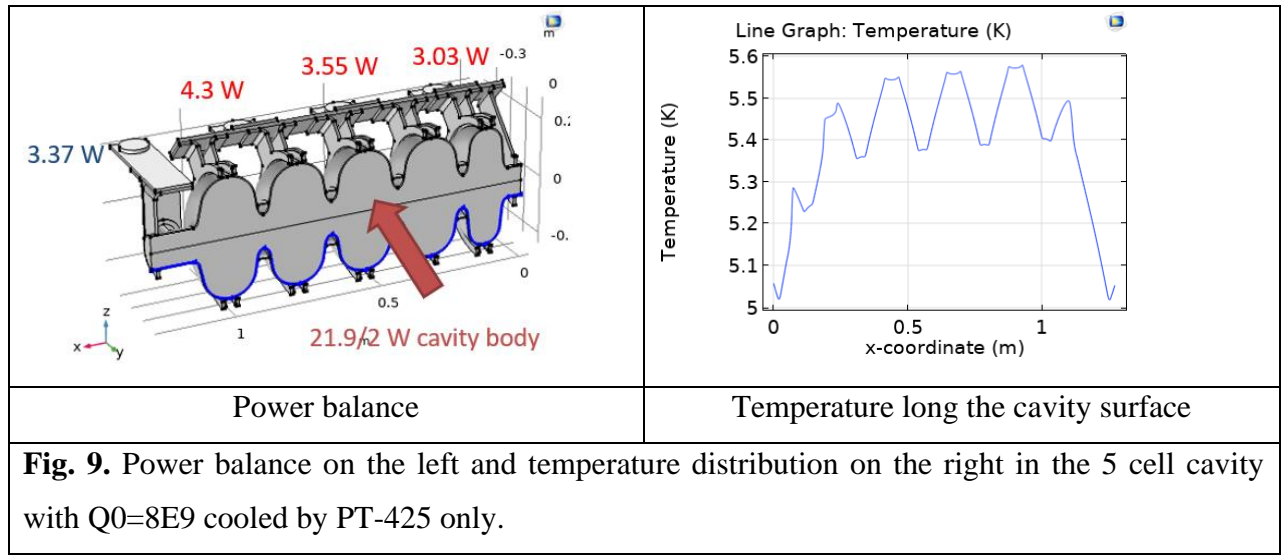

| Power balance | Temperature long the cavity surface |

**Fig. 9.** Power balance on the left and temperature distribution on the right in the 5 cell cavity with $Q_0$=8E9 cooled by PT-425 only.

Different cryoccolers should stabilize at different temperatures if the power coming to the cryocoolers is not the same. In order to find a correct solution, capacity map was used instead of the constant temperatures. Each cryocooler temperature was set up as a function of the power coming to a cryocooler accrding to the capacity map. Figs. 10 and 11 represent the case of cooling with both PT-420/PT-425 and PT-425 only correspondingly.

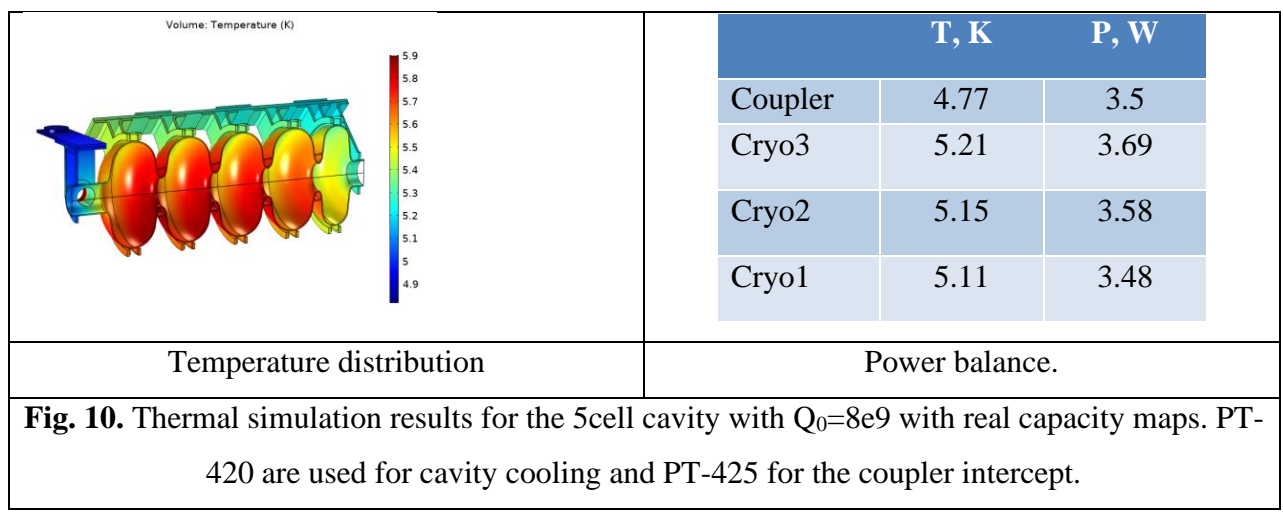

|  | T, K | P, W |
|---|---|---|
| Coupler | 4.77 | 3.5 |
| Cryo3 | 5.21 | 3.69 |
| Cryo2 | 5.15 | 3.58 |
| Cryo1 | 5.11 | 3.48 |

| Temperature distribution | Power balance. |

**Fig. 10.** Thermal simulation results for the 5cell cavity with $Q_0$=8e9 with real capacity maps. PT-420 are used for cavity cooling and PT-425 for the coupler intercept.

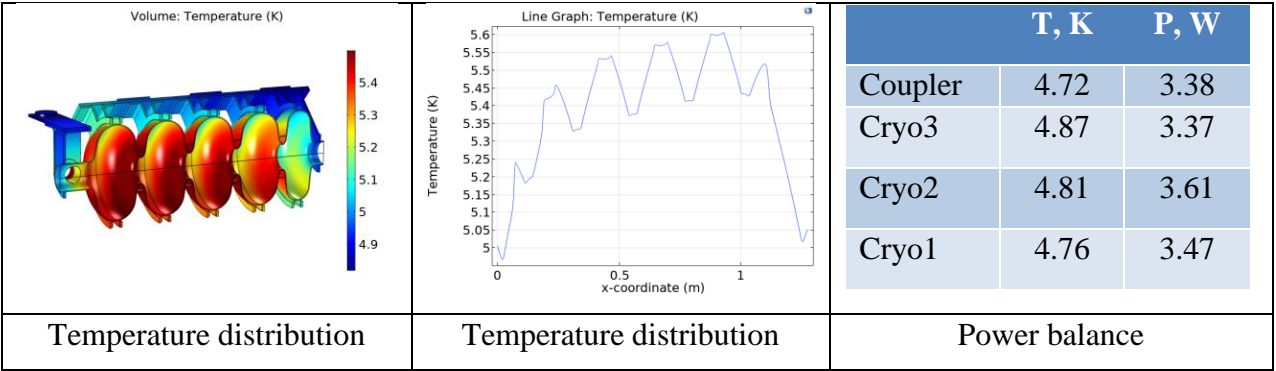

| | T, K | P, W |
|---|---|---|
| Coupler | 4.72 | 3.38 |
| Cryo3 | 4.87 | 3.37 |
| Cryo2 | 4.81 | 3.61 |
| Cryo1 | 4.76 | 3.47 |

| Temperature distribution | Temperature distribution | Power balance |

**Fig. 11.** Thermal simulation results for the 5cell cavity with $Q_0$=8e9 with real capacity map. Only PT-425 are used for cooling.

As one can see, the only-PT425 case has a lower temperature of the cavity. The power balances are also presented. Temperatures of the cryoccolers as well as the power coming to them are recorded. These numbers accurately correlerate with the capacity maps that were used. The power dissipated in the cavity is equal to the power taken by the cryocoolers. The use of the capacity map instead of the fixed temperatures of the cryocoolers provides a more realistic simulation. Otherwise an iterative simulation is required to match the heat coming to each cryocooler at certain tempeature with the capacity map and with the latest set-up it is done internall by Comsol.

## 4. The 2$^{nd}$ geometry case: bigger OD beam pipe.

### a. RF simulations

The cavity geoemtry was changed in order to increase the coupling with the fundamental power coupler. The last cell and the high energy side beam pipe were modified. The beam pipe was enlarged to increase the coupling, but that required the last cell tuning to keep the frequency and the field balance. The new geoemtry is presented in Fig. 12. Vacuum volume presented in this figure inherited the first cell geometry from the cavity with internal injection (OD=35 mm), which is included below to highlight the differences in geometries. The first cell input iris of 100 mm was used in the simulations whoch corresponds to the cavity with external injection.

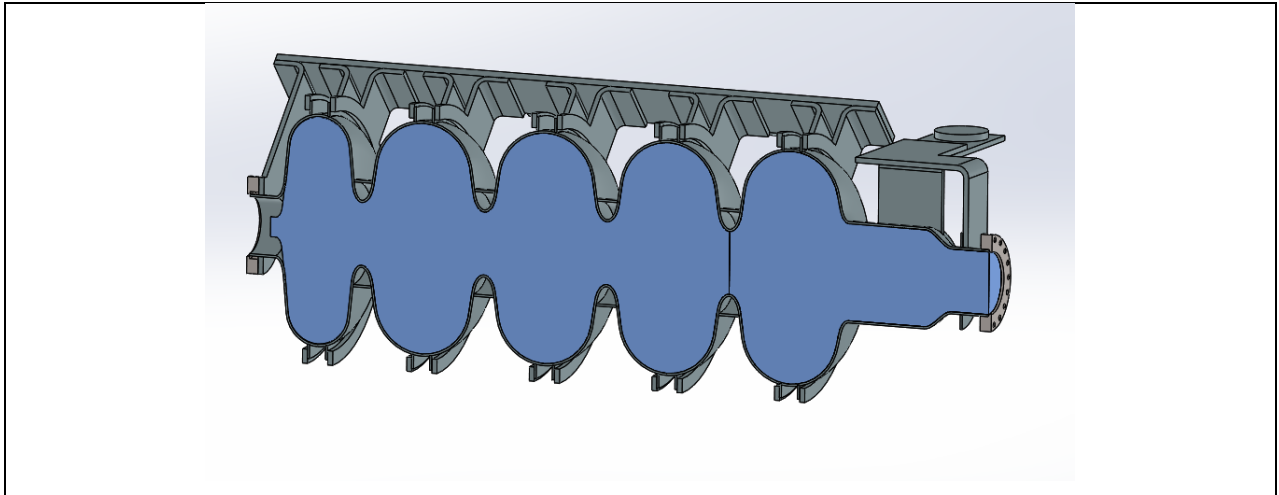
**Fig. 12.** A new geometry thermal and vacuum 3D model with 100 mm diameter inlet beam pipe.

RF simulations were done for obtaining new dynamic heat distribution in the new cavity. The results are presented in Fig. 13.

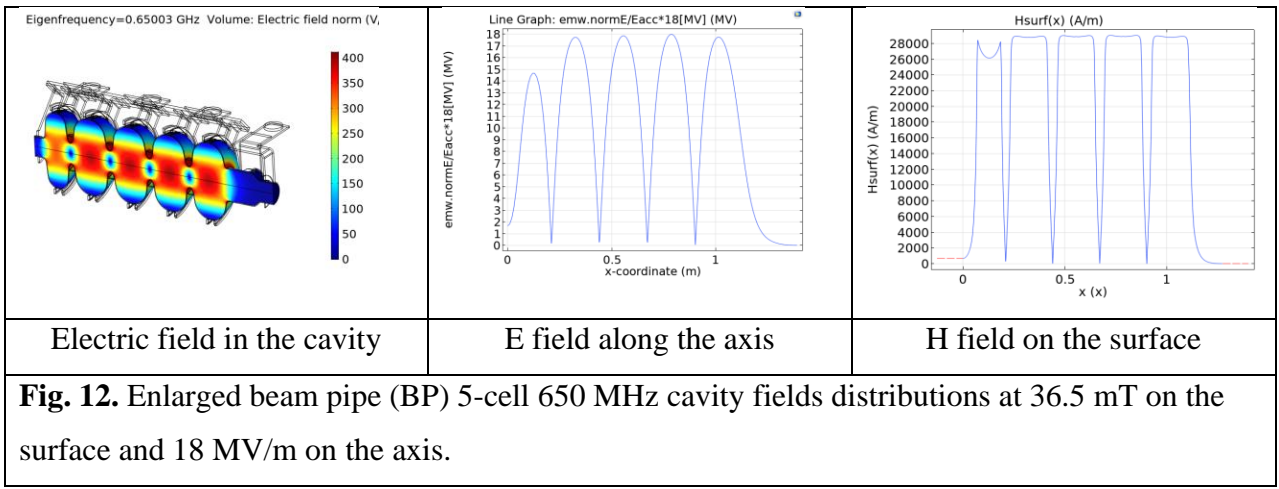

| Electric field in the cavity | E field along the axis | H field on the surface |

**Fig. 12.** Enlarged beam pipe (BP) 5-cell 650 MHz cavity fields distributions at 36.5 mT on the surface and 18 MV/m on the axis.

## b. Thermal simulation: $Q_0=1.5e10$, $Q_0=8e9$

Thermal simulations were done for several different cases: $Q_0=1.5e10$ and $Q_0=8e9$ cooled by PT420 and PT-425. The case of $Q_0=8e9$ was also simulated with PT-425 only. The results are presented below (Fig.13 and Table 3). The capacity maps of the cryocoolers were used in the simulations.

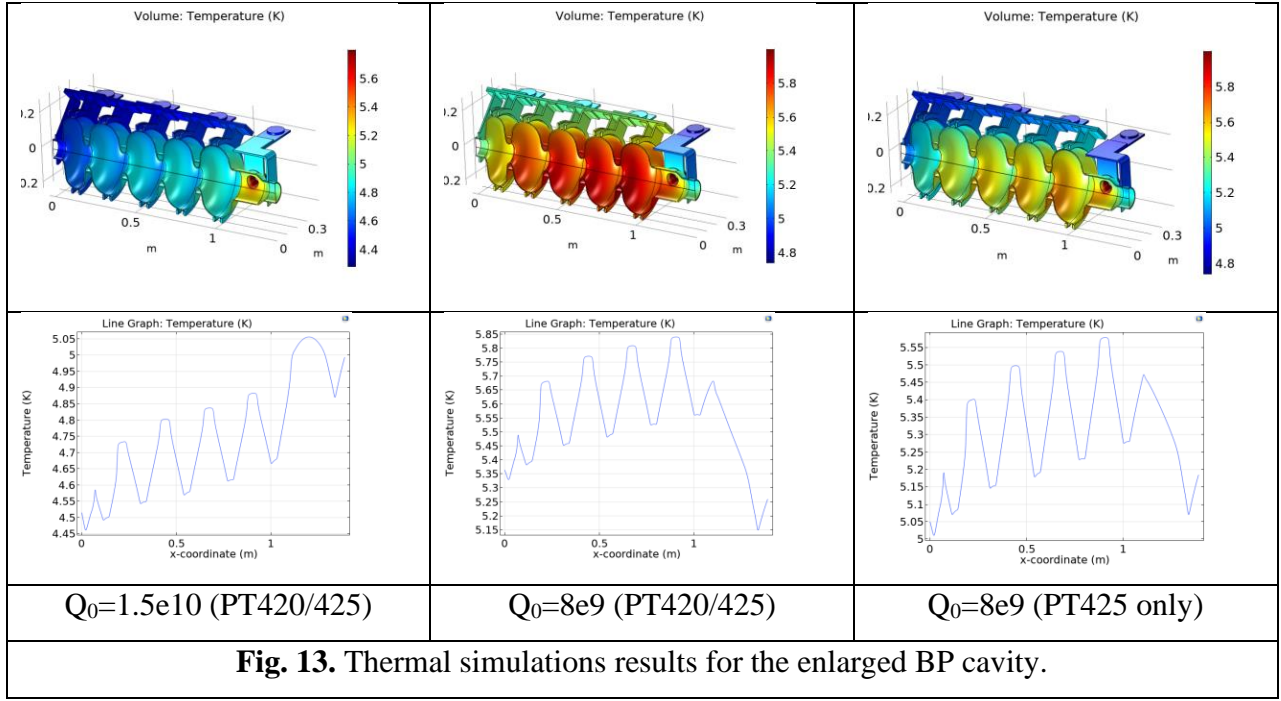

| | | |
|---|---|---|
| $Q_0$=1.5e10 (PT420/425) | $Q_0$=8e9 (PT420/425) | $Q_0$=8e9 (PT425 only) |

**Fig. 13.** Thermal simulations results for the enlarged BP cavity.

Table 3. Temperature and the heat flow to different cryocoolers

| | $Q_0$=1.5e10, 17.5 n$\Omega$ | | $Q_0$=8e9, 32.4 n$\Omega$ | | $Q_0$=8e9, 32.4 n$\Omega$ (all PT 425) | |
|---|---|---|---|---|---|---|
| | T, K | P, W | T, K | P, W | T, K | P, W |
| Coupler | 4.5 | 2.98 | 4.74 | 3.41 | 4.67 | 3.28 |
| Cryo3 | 4.38 | 2.22 | 5.23 | 3.73 | 4.90 | 3.82 |
| Cryo2 | 4.32 | 2.13 | 5.17 | 3.61 | 4.83 | 3.65 |
| Cryo1 | 4.28 | 2.07 | 5.12 | 3.51 | 4.78 | 3.51 |

As expected, the obtained results for the enlarged beam pipe have almost no difference with the previous results of the cavity with the regular beam pipe.

### c. Thermal simulation: non-constant Rs=R$_{BCS}$(T)+10/20/25/30 nOhm

An additional round of thermal simulations was done for the case of non-constant surface resistance, *i.e.*, BCS surface resistance was used as a function of temperature. PT420 and PT425 were used as described in the prior section. Several cases were studied: Rs=R$_{BCS}$(T)+10/20/25/30 nOhm. No convergence was found for the case of residual resistance of 30 nOhm, indicating that the dissipated power exceeds the cryocooler capacity. The cases that attained convergence are presented in Fig. 14 and Table 4.

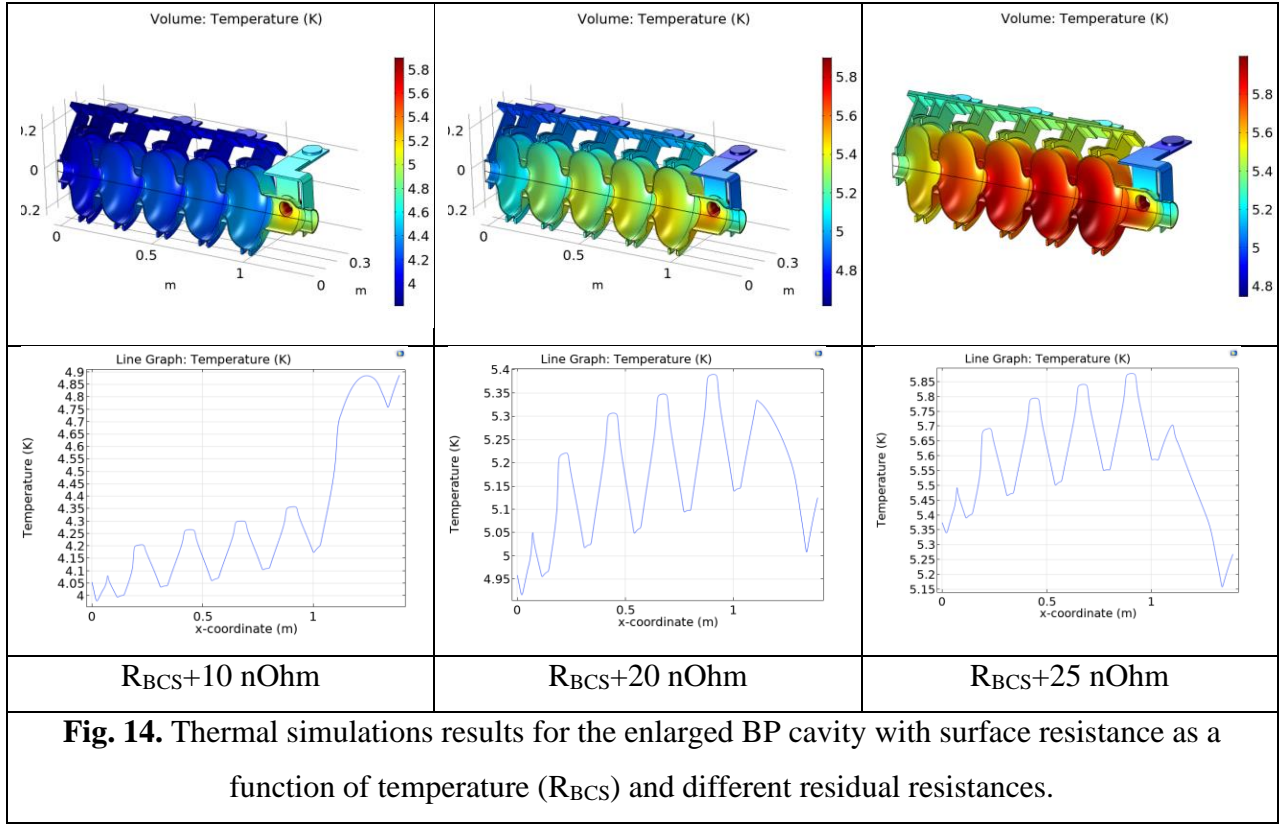

**Fig. 14.** Thermal simulations results for the enlarged BP cavity with surface resistance as a function of temperature ($R_{BCS}$) and different residual resistances.

The cases with $Q_0$=1.5e10 and $Q_0$=8e9 from the previous section roughly correspond to $R_{BCS}$+15 nOm and $R_{BCS}$+25 nOhm correspondingly.

Table 4. Temperature and the heat flow to different cryocoolers

|  | **Rbcs+10 nOhm** |  | **Rbcs+20 nOhm** |  | **Rbcs+25 nOhm** |  |
|---|---|---|---|---|---|---|
|  | T, K | P, W | T, K | P, W | T, K | P, W |
| Coupler | 4.4 | 2.81 | 4.62 | 3.19 | 4.75 | 3.44 |
| Cryo3 | 3.91 | 1.57 | 4.87 | 2.93 | 5.25 | 3.78 |
| Cryo2 | 3.85 | 1.49 | 4.77 | 2.80 | 5.19 | 3.65 |
| Cryo1 | 3.81 | 1.44 | 4.72 | 2.71 | 5.13 | 3.53 |

This section concludes that the cavity with the present configuration of thermal link and the cryocoolers must posses residual resistance below 30 nOhm for stable operation at 10 MeV energy gain.

## d. Thermal simulation: $R_S = R_{BCS}(T) + 10/20/25/30$ nOhm with a new thermal link.

A new thermal link design with less elements was simulated as well. The longitudinal bus now is connected directly to the cryocoolers, eliminating the final v-shaped small buses from the cryocoolers to the longitudinal bus. Non-constant surface resistance, *i.e.*, BCS surface resistance was used as a function of temperature. PT420 and PT425 were used as described in the prior section. Several cases were studied: $R_S = R_{BCS}(T) + 10/20/25/30$ nOhm. No convergence was found for the case of residual resistance of 30 nOhm, indicating that the dissipated power exceeds the cryocooler capacity. The cases with convergence are presented in Fig. 15 and Table 5. The new thermal link design helped to reduce the cavity temperature by 50 mK compared to the previous case. However, the thermal link and the number of cryocoolers are still insufficient for a cavity with residual resistance >30 nOhm.

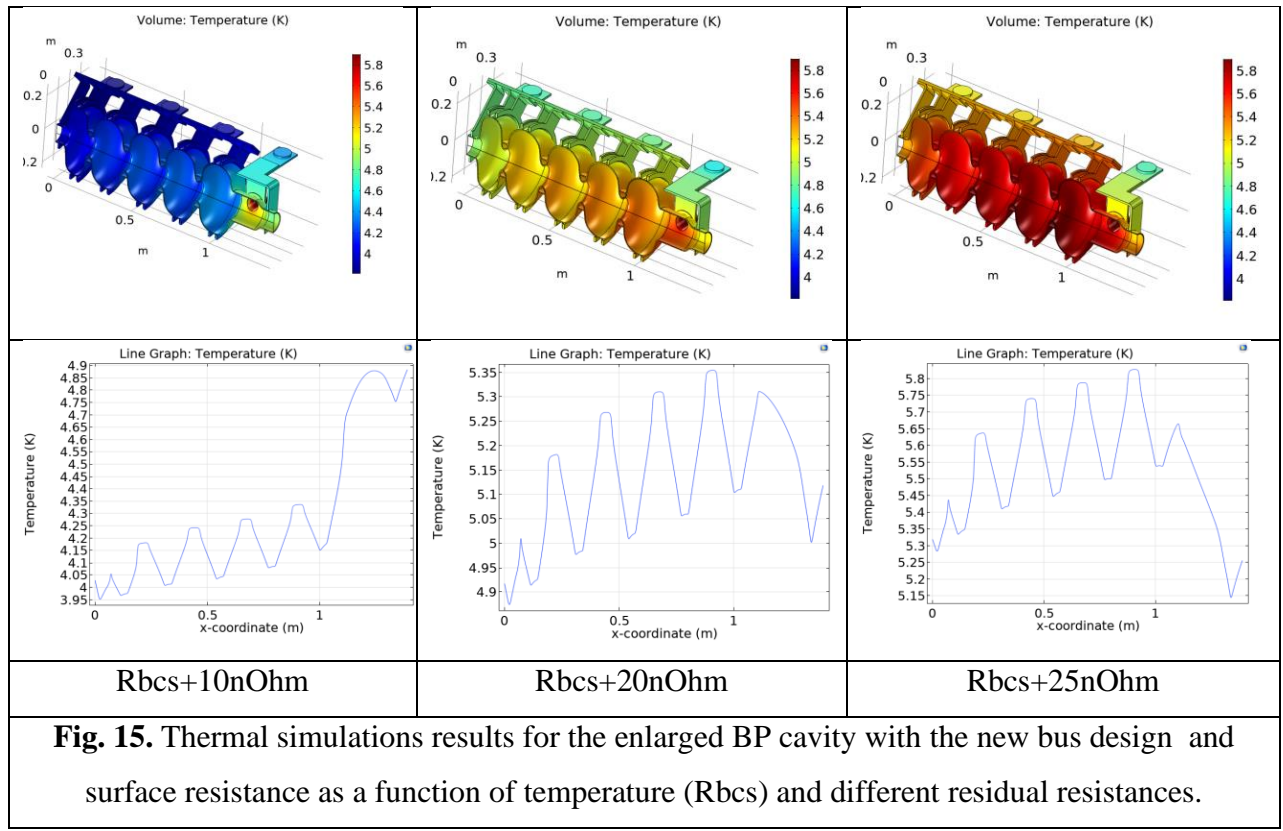

**Fig. 15.** Thermal simulations results for the enlarged BP cavity with the new bus design and surface resistance as a function of temperature (Rbcs) and different residual resistances.

Table 5. Temperature and the heat flow to different cryocoolers

|         | Rbcs+10 nOhm | | Rbcs+20 nOhm | | Rbcs+25 nOhm | |
|---------|------|------|------|------|------|------|
|         | T, K | P, W | T, K | P, W | T, K | P, W |
| Coupler | 4.4  | 2.80 | 4.62 | 3.17 | 4.74 | 3.41 |
| Cryo3   | 3.91 | 1.56 | 4.83 | 2.92 | 5.24 | 3.75 |
| Cryo2   | 3.85 | 1.48 | 4.77 | 2.79 | 5.17 | 3.61 |
| Cryo1   | 3.81 | 1.43 | 4.72 | 2.69 | 5.11 | 3.49 |

## 5. Conclusions

1. Two cases of 650 MHz 5-cell cavity with external injection of the beam were investigated: a standard beam pipe and an enlarged beam pipe for the increased coupling.
2. RF simulations were done to generate the dynamic heat load for the thermal module.
3. RF fields were scaled by surface magnetic field up to 36.5 mT which corresponded to 18 MV/m of electric field amplitude on the axis. Single particle tracking confirmed that the cavity could provide the 10 MeV energy gain for the 300 keV injected electrons at this field level.
4. Thermal simulations were performed for the cavity with the regular beampipe using the following conditions: nonlinear material properties, nonlinear contact resistance, and fixed cryocoolers temperatures estimated from the total dissipated heat. Two cases were investigated: $Q_0$=1.5e10 ($R_s$=10 nOhm), $Q_0$=8e9 ($R_s$=32 nOhm). Both cases yielded converged solutions. However, it was noticed that each cryocooler took out different amount of heat from the system. The simulation was refined with non-constant cryocooler temperatures as a function of heat coming to a cryocooler according to the capacity map. The refined simulation revealed different temperature distribution which described the system better and was used for the following simulations.

5. Thermal simulations were performed for the cavity with the enlarged beampipe. As expected, the obtained results for the enlarged beam pipe have almost no difference with the previous results of the cavity with the regular beam pipe.
6. As a next step of the model refinement, non-constant surface resistance was used as follows: $R_s=R_{BCS}(T)+R_{res}$(10/20/25/30 nOhm). Converged solutions were obtained for all cases except for the last one, meaning that the stable operation is not possible for the last case ($R_{res}$=30 nOhm). The cases with Q0=1.5e10 ($R_s$=10 nOhm) and Q0=8e9 ($R_s$=32 nOhm) from the previous section roughly correspond to $R_{BCS}(T)$+15 nOhm and $R_{BCS}(T)$+25 nOhm correspondingly.
7. Thermal simulations for the new thermal link design with less contacts were done for the cases with $R_s=R_{BCS}(T)$+10/20/25/30 nOhm. The new bus design helped to reduce the cavity temperature by 50 mK, however, the case with the residual resistance of $R_{res}$=30 nOhm was still unstable and did not converge.